\documentstyle[12pt]{article}
\begin{document}

\title{\sc  Proof of the Generalized Second Law for Quasistationary
Semiclassical Black Holes\thanks{Alberta-Thy-10-93, gr-qc/9302017}}
\vspace{.5cm}
\author{{\sc valery p. frolov}\thanks{On leave of absence from the
P. N. Lebedev Physical Institute, Leninsky prospect 53,
Moscow 117924, Russia} \thanks{Internet address:
frolov@phys.ualberta.ca} \vspace{.3cm} \\
{\small{and}}
\and
{\sc don n. page}\thanks{Internet address:
don@page.phys.ualberta.ca}\vspace{.3cm}
\\ {\small{\em CIAR  Cosmology Program}}
\\ {\small{\em Theoretical Physics Institute}}
\\ {\small{\em University of Alberta, Edmonton, Canada T6G 2J1}}}
\date{1993 Feb. 8; revised Aug. 17}
\vspace{.5cm}
\maketitle

 \vspace{1.5cm}
 \centerline{PACS Indices: 97.60.Lf; 05.30.-d; 04.20.Cv}

 \vspace{.4cm}

\begin{abstract}
A simple direct explicit proof of the generalized second law of
black hole thermodynamics is given for a quasistationary
semiclassical black hole.
\end{abstract}

\baselineskip=16.7pt
\vfill\eject
According to the thermodynamical analogy in black holes physics, the
entropy of a black hole in general relativity is defined as
$S_{BH}=A/4$,
where $A$ is the area of the black hole surface. Bekenstein \cite{1}
made an assumption that the generalized entropy $\widetilde{S}$,
i.e. the sum of the black hole entropy $S_{BH}$ and the entropy
$S_{rad}$ of the usual matter and gravitational radiation outside
a black hole ($\widetilde{S}=S_{BH} +S_{rad}$), never decreases.
This
assumption, known as the generalized second law, plays a fundamental
role in black hole physics. Though a simple explicit general proof
of this law has not been given until now,  concrete processes
were considered, and the validity of the generalized second law
for these concrete processes was verified.
(For a discussion of the generalized second law in black hole
physics,
see, e.g. \cite{3,2} and references therein.)

A special particular example is the case of an evaporating black
hole.  Quantum radiation created by a black hole carries entropy to
infinity, while the area of the black hole (and hence its entropy) is
decreasing.  The validity of the generalized second law for the
massless radiation evaporated by an uncharged, nonrotating
semiclassical black hole was almost proved by Zurek \cite{4}.
In particular, he estimated that
$R\equiv dS_{rad}/dS_{BH}\approx \frac{4}{3}$
for the evaporation of a black hole in vacuum.
(More accurate numerical
calculations of  the emission of neutrinos, photons, and
gravitons actually had given even higher values \cite{5,6}.)
The evaporation of charged black holes was discussed
by Schumacher \cite{7}.

Unruh and Wald \cite{8} stressed the importance of
the vacuum polarization and acceleration radiation effects for the
validity of the generalized second law.  More general arguments for
the validity of this law for slowly evolving black holes were given
by
Zurek and Thorne \cite{9} and by Thorne, Zurek, and Price \cite{10}
in the framework of the membrane paradigm.  The latter approach,
using the notion of the thermal atmosphere of a black hole,
looks quite general and  attractive, but it requires such an
operation
as the renormalization of the entropy density near a black hole. The
rigorous foundation of this operation is not completely clear
\cite{3},
though it can be argued that the renormalization should not affect
the derivation for quasistationary changes \cite{11}.

The aim of this paper is to present a simple explicit proof of the
generalized second law for quasistationary changes of a generic
charged rotating black hole emitting, absorbing, and scattering any
sort of radiation in the Hawking semiclassical formalism (quantum
radiation fields in the classical spacetime background of a black
hole whose conserved quantities change by the
expectation values of the flux of radiation out of or into it).
This proof may be considered to be a mathematical fleshing out of
some of the verbal arguments of Zurek, Thorne, and Price \cite{9,10}.

A quasistationary black hole may be considered to emit a density
matrix $\rho_0$ of thermal radiation \cite{12} in {\it UP} modes
\cite{10}.
Suppose there is also radiation with density matrix $\rho_1$
incident on the hole from far away (e.g., past null infinity, ${\cal
I}^-$)
in the corresponding {\it IN} modes
(i.e., incoming modes which are of positive frequency at ${\cal
I}^-$).
We use the semiclassical approximation and assume
that the radiation in these two sets of modes will be quantum
mechanically uncorrelated, giving an initial product state
\begin{equation}
\rho_{initial}\equiv\rho_{01}=\rho_0\otimes\rho_1. 
\end{equation}

This assumption is natural for an eternal black hole.
For it, the {\it UP} modes (defined
to be of positive frequencies with respect
to the Killing vector null at and tangent to the past horizon, $H^-$)
vanish at ${\cal I}^-$, whereas the {\it IN} modes
vanish at $H^-$, and ${\cal I}^-$ and $H^-$
are causally disconnected.

In the case in which the black hole arises from gravitational
collapse
and becomes quasistationary, the {\it UP} modes are defined
to be the same in the future stationary region as the {\it UP} modes
of the eternal black hole with the same future stationary region.
They are nonvanishing at ${\cal I}^-$
at the advanced time at which the black hole forms.
One might therefore worry that these modes
in principle could be correlated with the {\it IN} modes
which come from ${\cal I}^-$ at much later advanced time.
However, after the hole has become quasistationary,
the relevant {\it UP} modes trace back to such high energy modes
at ${\cal I}^-$ that the state in those modes must be
extremely close to the vacuum there.  Thus, in our quasistationary
approximation, they will have totally negligible
correlations with the {\it IN} modes coming in much later in advanced
time.  That is why, for the
physics of the quasistationary region at late time, both of the
pictures (eternal black hole and black hole arising from collapse)
give very nearly the same results.  For concreteness,
we shall use the eternal black hole picture in the following
discussion.

After the initial state $\rho_{01}$ interacts with the classical
angular
momentum and curvature barrier separating the horizon from infinity,
and possibly interacts with itself as well, it will have evolved
unitarily into a (generally) correlated final state
\begin{equation}
\rho_{final}\equiv\rho_{23}\ne\rho_2\otimes\rho_3,
\end{equation}
where
\begin{equation}
\rho_2=\mathop{\hbox{tr}}_3\rho_{23}
\end{equation}
is the density matrix of the radiation in the {\it OUT} modes
escaping to future null infinity ${\cal I}^+$, and
\begin{equation}
\rho_3=\mathop{\hbox{tr}}_2\rho_{23}
\end{equation}
is that in the {\it DOWN} modes that are swallowed by the future
horizon $H^+$.

The entropy of each of these states is
\begin{equation}
S_i = -\mathop{\hbox{tr}}(\rho_i\ln \rho_i).
\end{equation}
Because the evolution from $\rho_{01}$ to $\rho_{23}$ is unitary,
$S_{23} = S_{01}$.  Furthermore, since
$\rho_{01}$ is uncorrelated but $\rho_{23}$ is generically
partially correlated, the
entropies of these states obey the inequality \cite{Ar}
\begin{equation}
S_2+S_3\ge S_{23}=S_{01}=S_0+S_1.
\end{equation}

The first law of black hole physics \cite{13} for a black hole of
mass
$M,$ angular momentum $J,$ charge $Q$ (its conserved extensive
quantities), and Hawking temperature $T_{BH},$
angular velocity $\Omega,$ and electrostatic potential $\Phi$
(its thermodynamically conjugate intensive quantities) implies that
\begin{equation}
dS_{BH}=T_{BH}^{-1}(dM-\Omega dJ-\Phi dQ)=T^{-1} dE,
\end{equation}
where $T$ and $E$ are the local temperature and energy as measured by
a fiducial observer  (FIDO) \cite{10} corotating with the hole near
the
horizon. If $E_0$ and $E_3$ are the expectation values of the local
energies of the emitted state $\rho_0$ and the absorbed state
$\rho_3,$
respectively, then the semiclassical approximation, combined with the
first law, says that
\begin{equation}
\Delta S_{BH}=T^{-1}(E_3-E_0),
\end{equation}
assuming that the changes to the hole are sufficiently small that $T$
stays nearly constant throughout the process (the quasistationary
approximation).

Equation (8) and inequality (6) imply that the change in the
generalized entropy is
\begin{eqnarray}
\Delta\widetilde S &=&\Delta S_{BH}+\Delta S_{rad}=
T^{-1}(E_3-E_0)+S_2-S_1 \nonumber\\
&\ge& (S_0-T^{-1} E_0)-(S_3-T^{-1} E_3).
\end{eqnarray}
Now for fixed $T$ and for equivalent quantum systems (as the {\it UP}
modes of $\rho_0$ and the corresponding {\it DOWN} modes of
$\rho_3$ are by $CPT$ reversal, for a quasistationary black hole),
$S-T^{-1} E$ is a Massieu function \cite{14,15} (essentially the
negative
of the local free energy divided by the temperature) which is
maximized
for the thermal state $\rho_0.$ Therefore,
\begin{equation}
\Delta\widetilde S\ge 0, 
\end{equation}
which is the generalized second law.
This is an explicit mathematical demonstration of what Zurek, Thorne,
and Price \cite{9,10} argued verbally, that the generalized second
law is
a special case of the ordinary second law, with the black hole as a
hot, rotating, charged body  that emits thermal radiation
uncorrelated with
what is incident upon it.

This proof applies to any emission, scattering, or absorption
process, even for interacting fields (e.g., for boxes lowered to mine
a black hole \cite{8}, considered as special {\it IN} states, not as
changes in the boundary conditions), so long as the semiclassical
approximation applies, and so long as the changes take place in a
quasistationary manner so that, for example, Eq.~(8) is valid.
One would conjecture
that the generalized second law applies also for rapid changes to a
black hole \cite{10}, but then $S_{BH},$ one-quarter of the horizon
area,
would depend upon the future evolution. One would presumably also
need to include matter near the hole in $S_{rad},$ but it is
problematic how to do that in a precise way without getting
divergences from infinitely short wavelength modes if there is to be
a sharp cutoff to exclude matter inside the hole \cite{16,17}. In a
quasistationary process, one can with negligible error allow enough
time for the modes to propagate far from the hole, where the states
$\rho_1$ and $\rho_2$ and their respective entropies can be
evaluated unambiguously.

On the other hand, it is a controversial open question whether the
generalized second law applies for the fine-grained radiation entropy
$S_{rad}=-\hbox{tr}(\rho\ln\rho)$ outside of the semiclassical
approximation \cite{18}.
There an enormous number of (individually very small) nonthermal
multimode correlations in $\rho_0,$ perhaps induced at least in part
by
the quantum back-reaction of the metric, might conceivably restore
the full information of the initial collapse state to the final state
of the radiation after the hole evaporates away \cite{19,20,21,22}.
In other words, one might start off with a pure initial state in
which
$\widetilde S=S_{rad}=0,$ evolve unitarily to form a black hole state
in
which both $S_{BH}$ and $S_{rad}$ are positive (from the
coarse-graining
of dividing the total pure state system into black hole and radiation
subsystems and then adding the separate entropy of each, ignoring
their mutual correlations), and then go back to another pure
radiation state with $\widetilde S=S_{rad}=0$ after the hole
disappears and there is no more coarse graining in the definition
of $\widetilde S.$

To illustrate the proof of the generalized second law above, consider
the case in which $\rho_0,\ \rho_1,\ \rho_2,$ and $\rho_3$ are
the respective density matrices for single {\it UP}, {\it IN}, {\it
OUT},
and {\it DOWN} modes that are corresponding in the following sense:
Each mode is that of the same free quantum field
(the same particle-antiparticle species, assumed to interact only
with the semiclassical gravitational and electromagnetic
fields of the black hole) and has the same helicity and the same
total and orbital angular momentum quantum
numbers $j$ and $l$.  The {\it IN} and {\it OUT} modes each have the
same positive frequency $\omega > 0$ at past and future null
infinities,
the same azimuthal angular momentum quantum number $m$, and
the same charge $q$ (i.e., excitations of both are either particles
or
antiparticles of the quantum field).  The {\it UP} and {\it DOWN}
modes each have the same positive frequency
\begin{equation}
\widetilde{\omega}
=|\omega-m\Omega-q\Phi|\equiv\eta(\omega-m\Omega-q\Phi)
=\bar{\omega}-\bar{m}\Omega-\bar{q}\Phi > 0
\end{equation}
with respect to the Killing vector null at and tangent to
the past and future horizons, the same frequency-at-infinity
$\bar{\omega}=\eta\omega$,
the same azimuthal angular momentum $\bar{m}=\eta m$,
and the same charge $\bar{q}=\eta q$, where
\begin{equation}
\eta = \mbox{sgn}(\omega-m\Omega-q\Phi).
\end{equation}

If $\eta=+1$,
any three of the four modes are linearly related.
For superradiant modes (bosonic modes with $\eta=-1$),
the {\it UP} and {\it DOWN} modes must be complex
conjugated before any three of the four modes are linearly
related.

The expected numbers of particles $n_i$     in the states
$\rho_i$ are related as follows:
\begin{equation}
n_0 -n_3 =\eta(n_2 -n_1 ).
\end{equation}
The change in the generalized entropy due to the absorption and
emission in these modes is, using the inequality~(6),
\begin{equation}
\Delta\widetilde S=S_2-S_1-x(n_0 -n_3 )\ge S_0-S_3-x(n_0-n_3),
\end{equation}
where $x=T_{BH}^{-1}\widetilde{\omega}
=T_{BH}^{-1}|\omega-m\Omega-q\Phi|.$

The matrix elements of $\rho_0$ in a
Fock number basis in the Hilbert space of {\it UP} particles are
\begin{equation}
\rho_{0nn^\prime}=(1-\epsilon e^{-x})^\epsilon e^{-nx}
\delta_{nn^\prime},
\end{equation}
diagonal with the Boltzmann distribution. Here $\epsilon$ is $+1$
for bosons and $-1$ for fermions.

Thus
\begin{eqnarray}
n_0 &=&\sum_{n=0}^\infty \rho_{0nn}n=(e^{x}-\epsilon)^{-1}, \\
S_0 &=&-\hbox{tr}(\rho_0\ln\rho_0)=-\sum_{n=0}^\infty \rho_{0nn}\ln
\rho_{0nn} \nonumber\\
&=&x(e^{x}-\epsilon)^{-1}-\epsilon\ln(1-\epsilon e^{-x}) \nonumber\\
&=&\epsilon(1+\epsilon n_0)\ln(1+\epsilon n_0)-n_0\ln n_0.
\end{eqnarray}

The entropy $S_3$ in the {\it DOWN} mode depends on $\rho_1$
as well as $\rho_0$,
but for a given $n_3,$ it is maximized by the thermal distribution
with respect to $\widetilde{\omega}$
giving that value of $n_3$ (i.e., not at the same temperature as
$\rho_0$ if $n_3\ne n_0),$ which by the formula analogous to Eq.~(17)
gives the upper limit
\begin{equation}
S_3\le \epsilon(1+\epsilon n_3)\,\ln(1+\epsilon n_3)- n_3\ln n_3.
\end{equation}
Combining Eqs.~(14), (17), and (18) then gives
\begin{equation}
\Delta\widetilde S\ge f_\epsilon(n_0,n_3)\equiv(\epsilon+n_3)\,
\ln\frac{1+\epsilon n_0}{1+\epsilon
n_3}-n_3\,\ln\frac{n_0}{n_3}.
\end{equation}
Clearly $f_\epsilon=0$ when $n_0=n_3,$ and
\begin{equation}
\frac{\partial f_\epsilon}{\partial n_0}
=\frac{n_0-n_3}{n_0(1+\epsilon n_0)}
\end{equation}
has the same sign as $n_0-n_3,$ so $f_\epsilon=0$ is a global
minimum.
Therefore, $\Delta\widetilde S\ge 0$ from the effect of each
mode that acts independently.
modes).

The above proof is valid for both nonsuperradiant and superradiant
modes. It is instructive to discuss and compare these cases in
more detail.
The Hawking emission formula \cite{12} says that
\begin{equation}
n_2=(1-\Gamma )n_1 +\Gamma \tilde{n}_0 ,
\end{equation}
\begin{equation}
\tilde{n}_0\equiv (e^{\eta x}-\epsilon )^{-1}
=\eta n_0+\frac{1}{2}\epsilon(\eta-1),
\end{equation}
where $1-\Gamma$ is the fraction of a classical incident wave flux
that
is reflected from the {\it IN} mode to the {\it OUT} mode by the
angular
momentum and curvature barrier around the hole.
$\Gamma$ and $\tilde{n}_0$ are both negative for superradiant modes
(bosonic modes with $\omega-m\Omega-q\Phi<0$, i.e.,
$\epsilon=+1$ and $\eta=-1$).

For modes with $\eta=+1$ ($\omega-m\Omega-q\Phi > 0$),
Eq. (22) gives $\tilde{n}_0=n_0$, so Eqs. (21) and (13) imply
\begin{equation}
n_2=\Gamma n_0+(1-\Gamma)n_1,\qquad n_3
=(1-\Gamma)n_0+\Gamma n_1,
\end{equation}
\begin{equation}
n_2+n_3=n_0+n_1.
\end{equation}
This is just what one would expect from a barrier with transmission
probability
$\Gamma$ for each {\it  UP} particle to become an {\it OUT} particle
and
for each {\it IN} particle to  become a {\it DOWN} particle.
Similarly, there
is a
reflection probability $1-\Gamma$ for each {\it IN} particle to
become  an
{\it OUT} particle  and for each  {\it UP} particle  to become a
{\it
DOWN} particle.

For modes with $\eta=-1$ ($\omega-m\Omega-q\Phi < 0$),
Eq. (22) gives $\tilde{n}_0=-n_0-\epsilon$, so Eqs. (21) and (13)
imply
\begin{equation}
n_2=n_1-\epsilon\Gamma(1+\epsilon n_0+\epsilon n_1),\qquad
n_3=n_0-\epsilon\Gamma(1+\epsilon n_0+\epsilon n_1),
\end{equation}
\begin{equation}
n_2-n_3=n_1-n_0.
\end{equation}
In other words, in the absence of initial particles
$(n_1=n_0=0),$ the barrier produces an expected number
$n_2 = n_3 = -\epsilon\Gamma$ of
{\it DOWN-OUT} pairs (which is positive for $\eta=-1$,
since then $\epsilon$ and $\Gamma$ have opposite signs),
with the $n_2$ particles or antiparticles exiting in the
{\it OUT} mode and the $n_3$ {\it DOWN} antiparticles
or particles, respectively, entering the hole \cite{anti}.
This pair-production interpretation occurs in the
the ``old'' \cite{10} ``near-horizon'' \cite{23} viewpoint and
convention
we are using for the {\it UP} and {\it DOWN} modes
(with positive local frequencies $\widetilde{\omega}>0$)
rather than the ``new'' \cite{10} ``distant-observer'' \cite{23}
viewpoint
and convention (which uses complex conjugate {\it UP} and
{\it DOWN} modes and so has negative local frequencies near the
horizon, $\widetilde{\omega}<0$, when $\eta=-1$).
In our convention any initial particles or antiparticles present
remain
on the same side of the barrier (as given by the terms in Eq. (25)
independent of $\Gamma$), but they also induce stimulated emission
(for bosons, $\epsilon=+1,$ in the terms proportional to $n_0+n_1)$
or suppression (for fermions, $\epsilon=-1$) by the Pauli exclusion
principle.  (It is interesting to note that for fixed
$\omega-m\Omega-q\Phi<0$ and fixed $\Gamma>0$,
the fermion emission $n_2$ at constant $n_1$ therefore {\it
decreases}
with increasing temperature $T_{BH}$, which increases $n_0$.)

It is possible also to give a simple
direct proof of the generalized second law (which does not use
the inequality (6)) for the particular case in which
the density matrix $\rho_1$ in the {\it IN}
mode is thermal (i.e., diagonal in the Fock number basis, with
entries in a geometric sequence), though not necessarily with the
same temperature as $\rho_0$. The corresponding {\it OUT} density
matrix $\rho_2$ in this case will also be
thermal, with its temperature determined by $n_2,$ which in turn is
given by Eq.~(21) \cite{2}. (D.~N.~P. remembers conjecturing this
around
1975 to R.~P. Feynman. After initial disbelief, Feynman wrote out a
one-page proof that night.) Then Eqs.~(13), (14), and the analogues
of (17) for $S_1$ and $S_2$, give
\begin{equation}
\Delta\widetilde S = S_2-S_1 -\eta x(n_2-n_1)=Q_\epsilon(n_1),
\end{equation}
where
\begin{equation}
Q_\epsilon(n_1)\equiv\epsilon\,\ln\frac{1+\epsilon n_2}{1+\epsilon
n_1}
+n_1\,\ln\frac{n_1(1+\epsilon\tilde{n}_0 )}{\tilde{n}_0 (1+\epsilon
n_1)}+
n_2\,\ln\frac{\tilde{n}_0(1+\epsilon
n_2)}{n_2(1+\epsilon\tilde{n}_0 )}.
\end{equation}
Here Eq. (22) is used to express $\eta x$ in terms of $\tilde{n}_0$.
Eq.~(21) should be used to evaluate $n_2(n_1).$
In this special case we have an explicit expression~(28) for the
change of the generalized entropy in terms of the particle number
$n_1$ at ${\cal I}^-$, with no need for the inequality~(6)
and a discussion of states near the horizon.

To prove $\Delta\widetilde S\ge 0$ directly for this case, we need to
show
that $Q_\epsilon(n_1)\ge0$ for any allowed value for $n_1$. First we
note
that the function obeys the relations
\begin{eqnarray}
Q_\epsilon(\tilde{n}_0 )
&=&\frac{dQ_\epsilon}{dn_1}\,(\tilde{n}_0 )=0, \\
\frac{d^2Q_\epsilon}{dn_1^2}\,(n_1)
&=&\frac{An_1+B}{n_1(1+\epsilon n_1)n_2(1+\epsilon n_2)}.
\end{eqnarray}
Here
\begin{equation}
A=(1-\Gamma)\Gamma(1+2\epsilon\tilde{n}_0)
   =(1-\Gamma)\Gamma\eta(1+2\epsilon n_0),
\qquad B=\Gamma\tilde{n}_0 (1+\epsilon\Gamma\tilde{n}_0). 
\end{equation}
The quantities
\begin{equation}
1-\Gamma=|R|^2,\qquad \Gamma\tilde{n}_0 =n_2(n_1=0)\equiv n_{vac},
\qquad 1+\epsilon\Gamma\tilde{n}_0 =1+\epsilon n_{vac} 
\end{equation}
are all nonnegative, so $B\ge0.$ For bosons, $A\ge0$ also, so for all
$n_1\ge0,\ d^2Q_\epsilon/dn_1^2\linebreak[2]\ge 0$ for them.
For fermions $(\epsilon=-1),\ A$ is negative if (and only if) $\eta <
0,$
but if it is, the fermionic restriction
$n_1\le 1$ implies that
\begin{equation}
An_1+B\ge A+B=\Gamma(1-\tilde{n}_0 )(1-\Gamma+\Gamma\tilde{n}_0 )\ge
0,
\end{equation}
so $d^2Q_\epsilon/dn_1^2\ge 0$ for them as well.

For a nonsuperradiant mode, for which  $\tilde{n}_0 \ge 0,$
Eq.~(29) and the concavity  of $Q_\epsilon(n_1)$ implies that
$Q_\epsilon\ge 0.$ The function $Q_\epsilon(n_1)$ reaches its minimum
$Q_\epsilon=0$ at the point $n_1=\tilde{n}_0,$ where Eq.~(21)
implies that $n_2=\tilde{n}_0$ as well. In other words,
absorption and emission of thermal states in a nonsuperradiant mode
cannot decrease the generalized entropy $\widetilde S.$
The generalized entropy is not changed if and only if there is
thermal equilibrium between these {\it IN} and {\it OUT} states.

For a superradiant mode, Eq.~(25)  implies that $n_2>n_1.$
Since the thermal entropy function~(17) grows with $n,$ Eq.~(27) with
$\eta x<0$ directly shows that $\Delta\widetilde S>0.$
In other words, any thermal state in a superradiant {\it IN} mode
always increases the generalized entropy.

Finally, consider the case of a Schwarzschild black hole surrounded
by thermal radiation at a slightly different temperature,
$T_{rad}=T_{BH}(1-\Delta)$.
Summing over all particle species, helicities, and angular momenta,
and integrating over all mode frequencies $ \omega=(8\pi M)^{-1} x $,
one gets a total entropy production per entropy decrease of the hole
of approximately
\begin{equation}
\frac{d\widetilde S}{-dS_{BH}}\simeq
\frac{\sum\int dx x^2
e^x(e^x-\epsilon)^{-2}\Gamma(1-\frac{1}{2}\Gamma)}
{\sum\int dx x^2 e^x(e^x-\epsilon)^{-2}\Gamma}\Delta 
\end{equation}
(assuming that $\Delta$ and the interaction between different
modes is small).
This calculation, analogous to that \cite{5,6} of the entropy
production
of radiation into vacuum, is a refinement of Eq. (11) of Zurek
\cite{4}.
He effectively took $\Gamma$ to be either 0 or 1 for each mode and so
got the crude result $\Delta/2$, though we see that the actual
result is between $\Delta/2$ and $\Delta$.  The latter is what we
would
get once we allow the emitted radiation to thermalize with the
surrounding radiation.  In any case, the entropy increases
(since $-dS_{BH}$ and $\Delta$ have the same sign) until the black
hole
reaches the temperature of the radiation, assuming the total energy
is constrained to make that possible \cite{24}.

To summarize, we have proved that any emission, absorption, and/or
scattering by a quasistationary semiclassical black hole cannot
decrease the generalized entropy $\widetilde S=S_{BH}+S_{rad}.$
The generalized entropy $\widetilde S$ remains constant
 only when the incident radiation
in the {\it IN} modes is in the same thermal state as what the black
hole emits in the {\it UP} modes.
This is only possible for a black hole with no superradiant modes.
Such a black hole must either be uncharged and nonrotating or be
surrounded by a container which suppresses superradiant modes.

\vspace{12pt}
{\bf Acknowledgements}:\ \ The authors wish to thank Werner Israel,
Kip Thorne, and Robert Wald for helpful discussions and comments.
This work was supported by the Danish Natural Science Research
Council,
grant 11-9524-1SE, and by the Natural Sciences and Engineering
Research Council of Canada.

 \vfill\eject


\begin{thebibliography}{99}
\bibitem{anti} That is, the excitations of the {\it DOWN} mode have
opposite values of the energy-at-infinity,
azimuthal angular momentum, and charge from those of
the excitations of the {\it OUT} mode.  Therefore, the two
excitations are the antiparticles of each other if particles
and antiparticles are distinct for the species in question.
If the species does not have distinct particles and antiparticles,
as is the case for photons, then both the {\it OUT} and {\it DOWN}
excitations can simply be considered as particles,
though they will still have opposite values of the
energy-at-infinity and azimuthal angular momentum.
\end{thebibliography}
\end{document}